\documentclass[prl,twocolumn]{revtex4}
\usepackage{amsmath}
\usepackage{amssymb}
\usepackage{graphics}

\newcommand\ket[1]{\left|#1\right>}
\newcommand\bra[1]{\left<#1\right|}
\newcommand\braket[2]{\left<#1\right|\left.\!\!#2\right>}
\newcommand\expect[1]{\langle#1\rangle}

\newcommand\xo{\hat x}
\newcommand\xv{\vec{x}}
\newcommand\po{\hat p}
\newcommand\ro{\hat\rho}
\newcommand\Ao{\hat A}
\newcommand\Ho{\hat H}
\newcommand\Io{\hat I}

\newcommand\Uo{\hat U}
\newcommand\Ko{\hat K}

\newcommand\om{\omega}
\newcommand\al{\alpha}

\newcommand\sio{\hat\sigma}
\newcommand\siov{\hat{\vec{\sigma}}}
\newcommand\pov{\hat{\vec{p}}}

\newcommand\dg{^\dagger}

\newcommand\e{\mathrm{e}}

\newcommand\Tc{\mathcal{T}}

\newcommand\half{\frac{1}{2}}

\begin{document}
\title{Determination of the stationary basis from protective measurement on a single system}   
\author{Lajos Di\'osi}
\affiliation{Wigner Research Center for Physics\\
H-1525 Budapest 114. P.O.Box 49, Hungary}
\date{\today}
\begin{abstract}
We generalize protective measurement for protective joint measurement
of several observables. The merit of joint protective measurement is
the determination of the eigenstates of an unknown Hamiltonian rather
than the determination of features of an unknown quantum state.
As an example, we precisely determine the two eigenstates of an unknown
Hamiltonian by a single joint protective measurement of the
three Pauli matrices on a qubit state.  
\end{abstract}
\maketitle

\section{Introduction}
Protective measurement is one of the unexpected consequences of the strange structure 
of quantum mechanics. According to general wisdom, we cannot gain information on 
the unknown state $\ro$ of a single quantum system unless we distort the state itself.
In particular, we cannot learn the unknown state of a single system whatever test
we apply to it. It came as a surprise that in weak measurements \cite{AAV88} the 
expectation value $\expect{\Ao}$ of an observable $\Ao$ can be tested on
a large ensemble of identically prepared unknown states in such a way that 
the distortions per single systems stay arbitrarily small (cf. \cite{Dio06}, too). 
An indirectly related surprise came with the so-called protective measurements 
\cite{AhaVai93,AhaAnaVai93,AhaAnaVai96} capable to test $\expect{\Ao}$ at least in an 
unknown eigenstate of the Hamiltonian $\Ho$ at arbitrarily small distortion of the 
state itself. Interesting debates followed the proposal as to the merit of protective 
measurement in the interpretation of the wave function of a single system instead of 
a statistical ensemble (cf., e.g., \cite{Gao13} and references therein).    

My work investigates an alternative merit of protective measurement. 
First I construct joint protective measurements of several observables $\Ao_1,\Ao_2,\dots$ 
and re-state the original equations for them in a general form. Then I show that the 
straightforward task that a single joint protective measurement solves on a single system is the   
determination of the eigenstates of an otherwise unknown Hamiltonian. 

\section{Joint protective measurement of several observables}
\label{S2}
Consider a single quantum system in state $\ro$, and suppose that its Hamiltonian has
discrete non-degenerate spectrum $\om_1,\om_2,\dots$:
\begin{equation}\label{H}
\Ho=\sum_n \om_n\ket{n}\bra{n},
\end{equation}
with the eigenstates $\ket{n}$. 
Consider a set of Hermitian observables $\Ao_1,\Ao_2,\dots$. For later convenience,
introduce their expectation values in the stationary eigenstates:
\begin{equation}\label{Ann}
\expect{\Ao_\al}_n=\bra{n}\Ao_\al\ket{n},~~~\al=1,2,\dots
\end{equation}
To simultaneously measure the observables $\Ao_1,\Ao_2,\dots$, we use von Neumann
detectors with the canonical variables $(\xo_1,\po_1),(\xo_1,\po_1),\dots$, with vanishing
Hamiltonians. We prepare the detectors in state $\ro_D$ initially, such that the 
pointer variables $\xo_1,\xo_2,\dots$ are of zero means and of small dispersions $\delta x_1,\delta x_2,\dots$, respectively.
The conditions
\begin{equation}\label{deltaxAnm}
0<\delta x_\al\ll\vert\expect{\Ao_\al}_n-\expect{\Ao_\al}_m\vert
\end{equation}
must be satisfied for as many pairs $n\neq m$ as possible for each detector $\al=1,2,\dots$, to ensure 
that a maximum set of $\expect{\Ao_\al}_1,\expect{\Ao_\al}_2,\dots$ be distinguished by the detectors.
Now we introduce the
usual coupling $\Ko=\sum_\al\po_\al\Ao_\al/T$ between the observables and the 
detectors, respectively, with the factor $1/T$ where $T$ is the duration of the protective measurement.  

Let us evaluate the composite unitary dynamics  in interaction picture.
The observables and the coupling become time-dependent:
\begin{eqnarray}
\Ao_\al(t)&=&\e^{it\Ho}\Ao_\al\e^{it\Ho}\label{A_t},\\
\Ko(t)&=&\frac{1}{T}\sum_\al\po_\al\Ao_\al(t)\label{K_t}.
\end{eqnarray}
The unitary transformation after time $T$ reads  
\begin{eqnarray}\label{U_T}
\Uo_T&=&\Tc\exp\left(-i\!\!\int_0^T\!\!\!\!\Ko(t)dt\right)\nonumber\\
     &=&\Tc\exp\left(-i\sum_\al\po_\al\!\!\int_0^T\!\!\!\!\Ao_\al(t)\frac{dt}{T}\right)
\end{eqnarray}
where $\Tc$ stands for time-ordering. Inserting 
\begin{equation}\label{A_t_nm}
\Ao_\al(t)=\sum_{n,m}\e^{i(\om_n-\om_m)}\ket{n}\bra{n}\Ao_\al\ket{m}\bra{m},
\end{equation}
we find the contribution of the off-diagonal elements become  
heavily suppressed when $T\vert\om_n-\om_m\vert\gg1$ is satisfied for
all $n\neq m$. The ideal protective measurement requires $T=\infty$,
the corresponding unitary dynamics contains the contribution of
diagonal elements (\ref{Ann}) only:
\begin{equation}\label{U_infty}
\Uo_\infty=\sum_n\exp\left(-i\sum_\al\po_\al\expect{\Ao_\al}_n\right)\ket{n}\bra{n}.
\end{equation}
Observe that the eigenvalues $\omega_n$ of the Hamiltonian play no role, only 
the eigenstates $\ket{n}$ do.
The dynamics of joint protective measurement of (a finite number of) observables $\Ao_1,\Ao_2,\dots$ 
is captured by $\Uo_\infty$. It will entangle the system with the detectors in such a way that
the pointer variables $\xo_1,\xo_2,\dots$ get shifted by the expectation values of $\expect{\Ao_1}_n,\expect{\Ao_2}_n,\dots,$ 
taken in each eigenstate $\ket{n}$ in turn. The readout of the detectors will obtain the outcomes 
\begin{equation}\label{readout}
x_1=\expect{\Ao_1}_n\pm\delta x_1,~~x_2=\expect{\Ao_2}_n\pm\delta x_2,~~\dots
\end{equation}
with probability $\vert\braket{n}{\psi}\vert^2$. The terms $\pm\delta x_1,\pm\delta x_2,\dots$ indicate
statistical errors. The above outcomes mean that we have occasionally, 
i.e.: whenever the thresholds (\ref{deltaxAnm}) disclose the ambiguity of $n$, collapsed the state $\ro$ 
of the system into $\ket{n}\bra{n}$ and we have precisely (i.e.: at arbitrary small errors) measured 
the expectation values of $\Ao_1,\Ao_2,\dots$ in the stationary state $\ket{n}$ of $\Ho$.

Let us test the above dynamics on the uncorrelated pure initial state $\ket{\psi_D}\ket{\psi}$ 
of the system+detectors compound:
\begin{equation}\label{Upsi}
\ket{\psi_D}\ket{\psi}\longrightarrow\Uo_\infty\ket{\psi_D}\ket{\psi}.
\end{equation}
Let us introduce the wave function $\psi_D(x_1,x_2,\dots)$ of the detectors.
If we substitute (\ref{U_infty}) and multiply both sides of (\ref{Upsi}) by $\bra{x_1,x_2,\dots}$, we get
\begin{align}\label{mapro_M}
\psi_D&(x_1,x_2,\dots)\ket{\psi}\longrightarrow\nonumber\\
&\sum_n\psi_D(x_1-\expect{\Ao_1}_n,x_2-\expect{\Ao_2}_n,\dots)\ket{n}\braket{n}{\psi}.
\end{align}
This shows that, under the conditions (\ref{deltaxAnm}) on the initial wave function $\psi_D(x_1,x_2,\dots)$ of the detectors,
the state on the r.h.s. prepares the von Neumann measurement of $n$ and $\expect{\Ao_1}_n,\expect{\Ao_2}_n,\dots$.
In particular, the initial probability density $P(x_1,x_2,\dots)=\vert\psi_D(x_1,x_2,\dots)\vert^2$ 
changes like this:
\begin{align}\label{shiftP}
P(x_1,x_2,\dots)&\longrightarrow\\ 
&\sum_n\vert\braket{n}{\psi}\vert^2 P(x_1-\expect{\Ao_1}_n,x_2-\expect{\Ao_2}_n,\dots)\nonumber.
\end{align}
Formally, this expression is the statistical mixture corresponding to a von Neumann projective measurement of the stationary
basis resulting in the outcome $n$ with probability $\vert\braket{n}{\psi}\vert^2$.
In each term the initial positions of the pointers get shifted by the expectation values of
the corresponding observables in the given post-measurement eigenstate $\ket{n}$.  
The eigenvalues $\om_n$ themselves do not appear in the result, since they already canceled from
the unitary dynamics $\Uo_\infty$, as we observed before.

\section{Protective measurement of the stationary basis}
\label{S3}
We are going to show that a single joint protective measurement determines the eigenstates
of the unknown Hamiltonian. Our example is a single qubit in an unknown initial state
\begin{equation}\label{ro2}
\ro=\half(\Io+\vec{s}\;\siov),~~~~\vert\vec{s}\vert\leq1, 
\end{equation}
with unknown spatial polarization vector $\vec{s}$. Unknown is the Hamiltonian as well:
\begin{equation}\label{H2}
\Ho=\Omega\vec{e}\;\siov,~~~~\vert\vec{e}\vert=1,
\end{equation}
with unknown strength $\Omega$ and unknown direction $\vec{e}$ of the external 'magnetic' field.
The Hamiltonian has two unknown eigenvalues $\pm\Omega$ and eigenstates $\ket{\pm}$:
\begin{equation}\label{ketpm}
  \Ho=\Omega\ket{+}\!\!\bra{+}-\Omega\ket{+}\!\!\bra{+}
\equiv\Omega\frac{\Io+\vec{e}\;\siov}{2}
     -\Omega\frac{\Io-\vec{e}\;\siov}{2}.
\end{equation}
Now we prepare three von Neumann detectors and couple them to the three qubit observables
$\Ao_\al=\sio_\al$, for $\al=1,2,3$, respectively. Their joint protective measurement
is described by the unitary operator (\ref{U_infty}): 
\begin{equation}\label{U2_infty}
\Uo_\infty=\exp\left(-i\pov\expect{\siov}_+\right)\!\!\ket{+}\!\!\bra{+}
         +\exp\left(-i\pov\expect{\siov}_-\right)\!\!\ket{-}\!\!\bra{-}.
\end{equation}
With $\expect{\siov}_\pm=\bra{+}\siov\ket{+}=\pm\vec{e}$, the coupling shows
the following simple dependence on the unknown parameter $\vec{e}$ of $\Ho$:
\begin{equation}\label{U2_infty_e}
\Uo_\infty=\exp\left(-i\pov\;\vec{e}\right)\!\!\ket{+}\!\!\bra{+}
         +\exp\left(+i\pov\;\vec{e}\right)\!\!\ket{-}\!\!\bra{-}.
\end{equation}
This unitary operator acts on the initial uncorrelated state:
\begin{equation}\label{U2ro}
\ro_D\ro\longrightarrow\Uo_\infty\ro_D\ro\Uo_\infty\dg.
\end{equation}
Let the state $\ro_D$ be constrained by $\delta x_1,\delta x_2,\delta x_3\ll1$, cf. (\ref{deltaxAnm}).
Inserting (\ref{U2_infty_e}) and taking the diagonal matrix element $\bra{\vec{x}}\dots\ket{\vec{x}}$ 
of both sides, we get
the resulting change of the initial pointer statistics $P(\vec{x})=\bra{x}\ro_D\ket{x}$:  
\begin{equation}
P(\xv)\longrightarrow 
 \vert\!\!\bra{+}\ro\ket{+}\!\!\vert^2 P(\xv-\vec{e})
+\vert\!\!\bra{-}\ro\ket{-}\!\!\vert^2 P(\xv+\vec{e}).
\end{equation}
Expressing $\vert\bra{\pm}\ro\ket{\pm}\vert^2$ via (\ref{ro2}) and (\ref{ketpm}),
the final statistics of the pointers $x_1,x_2,x_3$ becomes 
\begin{equation}
 \frac{1+\vec{e}\;\vec{s}}{2}P(\xv-\vec{e})
+\frac{1-\vec{e}\;\vec{s}}{2}P(\xv+\vec{e}).
\end{equation}
If we read out the three detectors, the outcome is $\vec{x}\approx\pm\vec{e}$ with probability 
$(1\pm\vec{e}\;\vec{s})/2$, respectively. We have thus determined the spatial direction $\vec{e}$
of the external field at arbitrary high precision upto its sign though. The precision of the 
measured components $e_1,e_2,e_3$ is given respectively by the initial dispersions $\delta x_1,\delta x_2,\delta x_3\ll1$,
it does not depend on the initial state $\ro$ of the qubit. The strength $\Omega$ of the field remains unknown
while the obtained knowledge of $\pm\vec{e}$ means that we have precisely inferred the two stationary 
states $\ket{\pm}$. Our protective measurement collapses the system, exactly like an ideal von Neumann
measurement of $\Ho$ would do, into one of the two stationary states, just we cannot learn into which 
one of the two.

\section{Summary}
I have generalized the concept of protective measurement for joint protective measurement
of a (possibly finite) number of observables, determined the corresponding unitary operation
and its action on arbitrary uncorrelated initial state of the system and the detectors.  
I have shown that on a single qubit of unknown state and unknown Hamiltonian, the two
stationary states can be determined in a single joint protective measurement of the three Pauli matrices. 
The post-measurement state of the qubit is just like after a projective measurement of $\Ho$ it would be. 
This result may certainly be generalized for higher dimensional systems as well.
In fact, the full Hamiltonian can always be determined on a single system if, e.g., we perform a
suitable sequence of standard measurements. Yet the surprising feature of the joint protective
measurement is that the stationary states can be determined in a single step and in a transparent model.

This work was supported by the Hungarian Scientific Research Fund under Grant No. 75129
and the EU COST Action MP1006.

\end{document}